\documentclass[twocolumn,superscriptaddress,preprintnumbers,amsmath,amssymb,pra]{revtex4}
\usepackage{amssymb}
\usepackage{graphicx} 
\usepackage{color}
\usepackage[utf8]{inputenc}

\newcommand{\eps}[0]{\varepsilon}
\newcommand{\up}[0]{\uparrow}
\newcommand{\dn}[0]{\downarrow}
\newcommand{\lp}[0]{\left}
\newcommand{\rp}[0]{\right}
\newcommand{\s}[0]{\sigma}

\newcommand{\half}[0]{\frac{1}{2}}
\newcommand{\rarr}[0]{\rightarrow}
\newcommand{\rr}[0]{(\mathbf{r})}
\newcommand{\rrp}[0]{(\mathbf{r}')}
\newcommand{\darsec}[0]{\texttt{DARSEC}}

\newcommand{\rb}[0]{\mathbf{r}}

\begin{document}




\title{One-electron self-interaction and the asymptotics of the Kohn-Sham potential: \\ an impaired relation}

\author{Tobias Schmidt}
\affiliation{Theoretical Physics IV, University of Bayreuth, 95440 Bayreuth, Germany}

\author{Eli Kraisler}
\affiliation{Department of Materials and Interfaces, Weizmann Institute of Science, Rehovoth 76100, Israel}

\author{Leeor Kronik}
\affiliation{Department of Materials and Interfaces, Weizmann Institute of Science, Rehovoth 76100, Israel}

\author{Stephan K\"{u}mmel}
\affiliation{Theoretical Physics IV, University of Bayreuth, 95440 Bayreuth, Germany}


\begin{abstract}
One-electron self-interaction and an incorrect asymptotic behavior of the Kohn-Sham exchange-correlation potential are among the most prominent limitations of many present-day density functionals. 
However, a one-electron self-interaction-free energy does not necessarily lead to the correct long-range potential. This is here shown explicitly for local hybrid functionals. 
Furthermore, carefully studying the ratio of the von Weizs\"acker kinetic energy density to the (positive) Kohn-Sham kinetic energy density, $\tau_\mathrm{W}/\tau$, reveals that this ratio, which frequently serves as an iso-orbital indicator and is used to eliminate one-electron self-interaction effects in meta-generalized-gradient approximations and local hybrid functionals, can fail to approach its expected value in the vicinity of orbital nodal planes.
This perspective article suggests that the nature and consequences of one-electron self-interaction and some of the strategies for its correction need to be reconsidered. 
\end{abstract}


\maketitle

\section{Density functional approximations and their Kohn-Sham potentials}\label{sec.intro}

During the past decades, Kohn-Sham density-functional theory (DFT)~\cite{Hohenberg1964,Kohn1965} evolved into a standard tool for electronic structure calculations of atoms, molecules and solids. The decisive quantity of DFT is the exchange-correlation (xc) energy functional, $E_\mathrm{xc}$, which contains all electronic interaction beyond the classical electrostatic Hartree contribution, $E_\mathrm{H}$. 
$E_\mathrm{xc}$ in practice has to be approximated, and the approximation used governs the accuracy of a DFT calculation \cite{Kurth2000,Primer}. It is one of the puzzles of DFT that explicit density functionals such as the generalized gradient approximations (GGAs) can predict binding energies and bond lengths of complex many-electron systems reliably, but make substantial errors in describing simple one-electron systems. 
The underlying problem is well known as the one-electron ``self-interaction problem'' \cite{PZ'81}: For the exact functional, $E_\mathrm{xc}+E_\mathrm{H}$ will vanish for any one-electron ground-state density because one electron does not interact with itself -- but most approximate functionals yield a spurious finite value for this case. Following Ref.\ \cite{PZ'81} a functional is considered to be one-electron self-interaction free if it fulfills the condition
\begin{equation}
 E_\mathrm{H}[n_{i\s}] + E_\mathrm{xc}[n_{i\s}] = 0, \label{eq.si-error}
\end{equation}
where $n_{i\s} = |\varphi_{i\s}\rr|^2$ designates a single spin-orbital density.

Self-interaction plays a decisive (although not the only) role in the (un)reliability of density functional theory calculations, and its consequences are particularly pronounced, e.g., in questions of orbital localization \cite{PZ'81,TSW93,temmermanSIC,engelEXXbands}, ionization processes \cite{,tongchu,chu_rev,URS00,telnov}, charge transfer \cite{sicprl,tdsicprl,sicinteger}, and for the interpretability of eigenvalues and orbitals, e.g., as photoemission observables \cite{casida_elm,chong,pohl,PRBRC09,ksvgks,dauthprl,klupfel2013,lkskbook}.

Many of these observables can also be directly related to properties of the Kohn-Sham exchange-correlation potential, which is defined as the functional derivative of the xc energy with respect to the 
ground-state density $n\rr$, i.e., $v_\mathrm{xc}\rr=\frac{\delta E_\mathrm{xc}[n]}{\delta n\rr}$.
It is generally expected that there is a close relation between freedom from self-interaction and xc potential features. The field-counteracting term that is important for obtaining correct response properties is one such feature \cite{G99,KKP04}. Another example, and probably the most prominent one, is the long-range asymptotic behavior of the xc potential \cite{Levy1984,Almbladh1985},
\begin{equation}
 v_\mathrm{xc}\rr \underset{|\rb|\rightarrow\infty}{\longrightarrow}-\frac{1}{r}. \label{eq.vxclimit}
\end{equation} 
(Hartree units are used here and throughout.)
In this perspective article we focus exclusively on Kohn-Sham theory, i.e., on a local multiplicative xc potential, as opposed to orbital-specific (non-multiplicative) potentials that arise in generalized Kohn-Sham theory \cite{SGVML96} and are used in the standard application of hybrid functionals. In the Kohn-Sham approach, the local xc potential models the interaction of one particle with all others and it therefore appears intuitively plausible that a functional that is not self-interaction-free cannot show the correct $ -1/r$ long-range asymptotic behavior: As one particle of a finite, overall electrically neutral systems ventures out to infinity, it will ``feel'' the hole of charge $1$ that it left behind in the total charge. This gives rise to the $ -1/r$ potential asymptotics (see, e.g., Ref.\ \cite{Primer}, p.\ 242 for a more detailed argument along these lines). However, a particle that spuriously self-interacts will ``feel'' itself, and thus not the proper hole. Consequently, the potential will not have the proper long-range decay.

The correct asymptotics of the xc potential has proven to be important for a variety of physical quantities. It plays a prominent role for obtaining stable anions in DFT, it leads to a Rydberg series in the Kohn-Sham eigenvalues and generally to unoccupied eigenvalues of improved interpretability, and as a consequence allows for improved accuracy in the prediction of various response properties \cite{Tozer1998,Tozer99,Casida2000,marques2001,tozershieldings}. The correct asymptotic behavior is also important for the ionization potential (IP) theorem~\cite{PPLB82,Levy1984,Almbladh1985,PerdewLevy97}, which states that the negative of the highest occupied Kohn-Sham eigenvalue $-\eps_\mathrm{ho}$ should correspond to the vertical IP, and for developing functionals that allow for approximately predicting IPs from ground-state eigenvalues \cite{tuningperspective,VermaBartlett2012}. 

There have been fruitful attempts to incorporate the correct behavior in the limit $|\rb|\rarr\infty$ directly into the xc potential~\cite{RvLBaerd94,TozerHandy,Becke2006,Cencek2013}, leading to improvements in the description of some of the aforementioned properties. However, since directly designed potential expressions are typically not functional derivatives of any energy functional, the use of such ``potential only'' approximations is necessarily limited, as discussed, e.g., in detail in Refs.\ \cite{stray,KAK09,karolewski2013}.

A functional that combines freedom from self-interaction and the correct asymptotics of the potential is exact exchange (EXX), being defined as the Fock integral evaluated using Kohn-Sham orbitals $\varphi_{i\s}\rr$, where $i$ labels orbitals in spin channel $\s$:
\begin{equation}
E_\mathrm{x}^\mathrm{ex}\rr=-\half\sum_{\substack{ i,j=1\\
\s=\up,\dn}}^{N_{\s}}\int\!\!\!\int
\frac{\varphi_{i\s}^*\rr\varphi_{j\s}\rr\varphi_{i\s}  \rrp\varphi_{
j\s}^*\rrp}{|\mathbf{r}-\mathbf{r}'|}\,\mathrm d^3r\mathrm d^3r' \, .
\label{eq.exx}
\end{equation}
Here, $N_{\s}$ is the number of electrons with spin $\s$. Treating exchange exactly with a local Kohn-Sham potential leads to a significant improvement in the quality of Kohn-Sham eigenvalues when comparing to (semi-)local functionals~\cite{Grabo1997,lhf,ksvgks}.
EXX also tends to increase Kohn-Sham gaps \cite{gorlingEXXgaps,engelEXXbands,makmal2009pp,engel2009pp,betzinger2013}, leads to a desired particle number discontinuity in static \cite{KLI92a} and time-dependent \cite{mundtprl} situations, and improves the description of charge transfer \cite{G99,KKP04}, dissociation \cite{makmal2011} and ionization processes \cite{mundtprl}.  

However, using bare EXX is known for its rather poor description of binding energies and structural properties (see, e.g., Refs.~\cite{Engel2000d,EngelDreizler2011}, and Ref.~\cite{Primer}, chapter 2).  
Adding a (semi-)local correlation term to EXX hardly improves the situation and typically leads to results that are inferior to the ones from (semi-)local functionals. The reason for this failure is the well-known incompatibility of the fully non-local Fock exchange with a purely (semi-)local correlation term~\cite{PerdewSchmidt2001}.

A class of approximations which has been designed to remedy this incompatibility is the one of local hybrid functionals \cite{Cruz1998,Jaramillo2003}, sometimes also called hyper-GGAs \cite{PerdewSchmidt2001}. 
Whereas global hybrid functionals \cite{Becke1993,Becke1993a,Perdew1996b,B3LYP_2,Adamo1999a,ErnScu} mix a constant, fixed fraction of Fock exchange with (semi-)local exchange and correlation,
local hybrids replace the fixed fraction by a density dependent local mixing function (LMF). Both types of hybrids originate from the concept of the coupling-constant integration, i.e., adiabatic connection scheme \cite{Becke1993,couplingconst}. Global hybrids are successful in modeling the coupling-constant averaged, integrated energy. Local hybrids can go one step further and aim to model the coupling-constant curve itself \cite{adiaconnect} instead of just the integral. Thus, in contrast to the global hybrid functionals which are used in practical applications of DFT and combine GGA components with about 25 \% of exact exchange, local hybrids can incorporate full exact exchange and can be fully one-electron self-interaction-free.

An early local hybrid with reduced one-electron self-interaction error showed promising results for dissociation curves and reaction barriers, but its accuracy for binding energies was limited \cite{Jaramillo2003}. A self-consistent implementation of a local hybrid functional was given in Ref.\ \cite{Arbuznikov2006}, and over the years several local hybrids were constructed, using different LMFs and (semi-)local exchange and correlation functionals \cite{Janesko2007,Arbuznikov2007,Bahmann07,Kaupp07,Perdew2008,haunschild2009,ArbuzBahmKaupp09,haunschild2010_2,Theilacker11}, 
striving to reach greater accuracy by refining the position-dependent mixing of nonlocal and local components. 
Many of these functionals rely on the concept of an iso-orbital indicator, i.e., a functional that allows one to distinguish regions of space in which the density is dominated by one orbital shape from regions of space where several orbitals of different shape contribute to the density. 
The most prominent iso-orbital indicator, which goes back to a long tradition of using kinetic energy densities in density functional construction \cite{becke85,dobson92,becke98}, is the ratio of the von Weizs\"acker kinetic energy density $\tau_\mathrm{W}$ to the positive (as opposed to other possible definitions, see, e.g., Ref.\ \cite{molphys03}) Kohn-Sham kinetic energy density $\tau$, discussed in detail below. 

By using full EXX and an iso-orbital indicator, local hybrids aim at being one electron self-interaction-free and producing a Kohn-Sham potential with the proper long-range asymptotic decay. They are a paradigm class of functionals designed for simultaneously curing both of these two prominent problems of (semi-)local density functionals. 
In the following, we therefore use the example of a local hybrid functional to shed light on the relation between a functional's self-interaction and its potential asymptotics, as well as the properties of the $\tau_\mathrm{W}/\tau$ indicator. We argue that quite generally a one-electron self-interaction-free energy does not guarantee the correct long-range potential, and that $\tau_\mathrm{W}/\tau$ loses its indicator ability in the vicinity of nodal planes of the highest-occupied molecular orbital (HOMO).

\section{Correlation compatible with exact exchange: the local hybrid approach}\label{sec.theory_LHF}

The xc energy functional can be written as 
\begin{equation}
 E_\mathrm{xc}[n]= \int n\rr\,e_\mathrm{xc}([n];\rb)\,\mathrm d^3r, \label{eq.xc_gen}
\end{equation}
with $e_\mathrm{xc}([n];\rb)$ denoting the xc energy density per particle. The definition of $e_\mathrm{xc}\rr$ is not unique and subject to a gauge-dependence~\cite{Perdew2008}. Yet, for local hybrid functionals it has become common to define this energy in the form
\begin{equation}
 e^\mathrm{lh}_\mathrm{xc}\rr = e_\mathrm{x}^\mathrm{ex}\rr + f\rr(e_\mathrm{x}^\mathrm{sl}\rr-e_\mathrm{x}^\mathrm{ex}\rr)+e_\mathrm{c}^\mathrm{sl}\rr \label{eq.lh_spec}.
\end{equation} 
 Here, $e_\mathrm{x}^\mathrm{ex}\rr$ marks the exchange energy density per particle corresponding to the EXX energy of Eq.~(\ref{eq.exx}). This nonlocal term is mixed with (semi-)local exchange and correlation energy densities $e_\mathrm{x}^\mathrm{sl}\rr$ and $e_\mathrm{c}^\mathrm{sl}\rr$, respectively. The position dependent mixing ratio $f\rr$, which is itself a density functional, marks the LMF.

Often, the LMF is designed in a way that aims at eliminating the one-electron self-interaction error of Eq.\ (\ref{eq.si-error}) that is inherent in most density functionals. 
An established method for reducing self-interaction effects is to detect regions of space where a single Kohn-Sham orbital shape dominates the density (``iso-orbital regions''), and then enforce Eq.\ (\ref{eq.si-error}) in these regions. 
One of the most popular \cite{pkzb,Jaramillo2003,molphys03,tpss,Arbuznikov2006,Bahmann07,Kaupp07,ArbuzBahmKaupp09,Theilacker11,iso} indicator functions for detecting iso-orbital regions is
\begin{equation}
 g\rr = \frac{\tau_W \rr}{\tau\rr}, \label{eq.iso-ind}
\end{equation}
where $\tau_W\rr = |\nabla n\rr|^2 / (8n\rr)$ denotes the von Weizs\"{a}cker kinetic energy density and
 $\tau\rr = \half \sum_\s \sum_{i=1}^{N_\s} |\nabla \varphi_{i\s}\rr|^2$ is the positive Kohn-Sham kinetic energy density.
In iso-orbital regions, $\tau\rr \rarr \tau_W\rr$ and therefore $g\rr \rarr 1$. In the case of a slowly varying density, $\tau_W\rr \rarr 0$ and, since $\tau\rr$ remains finite, $g\rr \rarr 0$. This indicator function is typically a decisive ingredient in the LMF, $f\rr$, of local hybrids. With its help one can construct $f\rr$ such that Eq.~(\ref{eq.lh_spec}) reduces to correct limiting cases, e.g.,
$e_\mathrm{x}^\mathrm{sl}\rr +e_\mathrm{c}^\mathrm{sl}\rr$ for slowly varying densities, and $e_\mathrm{x}^\mathrm{ex}\rr$ for single orbital regions. The latter case additionally requires that $e_\mathrm{c}^\mathrm{sl}\rr$ vanishes in single-orbital regions, a condition that we discuss below.

In the asymptotic limit, $|\mathbf{r}|\rarr\infty$, the xc energy density for a finite system should be dominated by $e_\mathrm{x}^\mathrm{ex}\rr$.
When $e_\mathrm{c}^\mathrm{sl}\rr$ vanishes sufficiently fast in the asymptotic region (a condition that is usually fulfilled), then
\begin{equation}
 \lim_{|\rb|\rightarrow\infty}f\rr = 0 \label{eq.f-asy}
\end{equation}
is the requirement that one aims at, because it leads to the correct asymptotic limit of the xc energy density per particle
\begin{equation}
 e_\mathrm{xc}^\mathrm{lh}\rr \sim e_\mathrm{x}^\mathrm{ex}\rr \underset{|\rb|\rightarrow\infty}{\longrightarrow} -\frac{1}{2r}\label{eq.exc-asy}.
\end{equation}
(Note the difference to the asymptotic limit of the xc potential, see Ref.\ \cite{RvLBaerd94}).

Since for a finite system each Kohn-Sham orbital decays exponentially with an exponent set by its eigenvalue \cite{kreibich}, the density is asymptotically dominated by the HOMO density, i.e., becomes of iso-orbital character. Therefore, $g\rr$ can be used in the construction of the LMF to realize Eq.\ (\ref{eq.exc-asy}).

Considerations of the type discussed above are inherent to many density functional constructions. As a particular example for a local hybrid functional we here use 
a recently proposed, physically motivated LMF~\cite{iso}, which reads
\begin{equation}
 f_{t}\rr = \frac{1 - \frac{\tau_W\rr}{\tau\rr} \zeta^2\rr}{1+ct^2\rr}. \label{eq.fiso}
\end{equation}
The function $g\rr$ in the numerator is multiplied by the squared spin polarization $\zeta\rr = (n_\up\rr - n_\dn\rr)/(n_\up\rr + n_\dn\rr)$, which lets the LMF not only identify iso-orbital regions, but also correctly distinguish between true one-orbital regions, and regions with two identical spin-orbitals. 
The function $g\rr$ is used in such a way that $f_t\rr$ vanishes for one-orbital regions, as required. The use of the reduced density gradient
\begin{equation}\label{eq.red_den_grad}
t^2\rr = \lp( \frac{\pi}{3} \rp)^{1/3} \frac{a_0}{16 \Phi^2(\zeta\rr)} \frac{|\nabla n\rr|^2}{n^{7/3}\rr},
\end{equation}
where $a_0$ is the Bohr radius and $\Phi(\zeta\rr) = \half \lp( (1+\zeta)^{2/3}+(1-\zeta)^{2/3}\rp)$, in the denominator of $f_{t}\rr$, ensures the correct
behavior of $E_\mathrm{xc}$ under uniform coordinate scaling  $\rb \rarr \gamma \rb$~\cite{Levy1985,Levy1991}. The density transforms as $n_\gamma\rr = \gamma^3 n(\gamma \rb)$ and as a consequence Eq.\ (\ref{eq.fiso}) uses full exact exchange in the sense of \cite{Perdew2008}
\begin{equation}\label{eq.fexx}
\lim_{\gamma\rightarrow\infty} \frac{E_\mathrm{xc}[n_\gamma]}{E^\mathrm{ex}_\mathrm{x}[n_\gamma]}=1.
\end{equation}
The function $t^2\rr$ is multiplied by a parameter $c$ that we cannot determine, at least presently, from fundamental constraints. 
It allows for adjustments in the functional ansatz. In the case of slowly varying densities, $f_t\rr \rarr1$ and
Eq.~(\ref{eq.lh_spec}) reduces to its purely (semi-)local components. 
As an aside we note that this LMF comprises the one of Ref.\ \cite{Jaramillo2003} as the special case $c=0$ and $\zeta\rr=1 \, \forall \, \rb$. We denote this case by $f_0\rr$, i.e.,  $f_0\rr=1 - \frac{\tau_W\rr}{\tau\rr}$.

For the semi-local exchange we use the LSDA~\cite{Parr1989}, i.e., $e^\mathrm{sl}_\mathrm{x}\rr=e^\mathrm{LSDA}_\mathrm{x}\rr$, whereas $e^\mathrm{sl}_\mathrm{c}\rr= \lp( 1 - \frac{\tau_W\rr}{\tau\rr} \zeta^2\rr \rp) e^\mathrm{LSDA}_\mathrm{c}\rr$.
The additional multiplication with the numerator of Eq.~(\ref{eq.fiso}) consistently reduces Eq.~(\ref{eq.lh_spec}) to pure EXX in the one-spin-orbital case, where $e^\mathrm{LSDA}_\mathrm{c}\rr$ alone does not vanish. 

The general questions that we discuss in this perspective article, i.e., whether there is a relation between self-interaction and the xc potential asymptotics and in how far the iso-orbital indicator $\tau_\mathrm{W}/\tau$ can be used to enforce freedom from self-interaction, can be scrutinzed with the local hybrid of Eq.~(\ref{eq.fiso}) as an instructive example.

\section{The Kohn-Sham exchange-correlation potential of local hybrid functionals}\label{sec.theory_asy}

In order to implement local hybrids self-consistently within the Kohn-Sham scheme, one has to find the local multiplicative xc potential corresponding to the energy of  Eqs.~(\ref{eq.xc_gen}) and (\ref{eq.lh_spec}). The fact that local hybrids
use EXX and typically also $\tau\rr$ makes them explicitly orbital-dependent.
Therefore, the local xc potential must be obtained from the optimized effective potential (OEP) equation (see, e.g., \cite{SH53,TS76,Sahni82,Grabo1997,EngelDreizler2011,Kummel2008}).
The computational effort can be reduced significantly by employing the approximation of Krieger, Li and Iafrate (KLI)~\cite{Li1992a}. For the local hybrid of Eq.~(\ref{eq.fiso}) it has been shown that the total energy $E_\mathrm{tot}$ and the highest occupied Kohn-Sham eigenvalue $\eps_\mathrm{ho}$ obtained with the KLI approximation agree quite well with the ones from the full OEP \cite{iso}. Furthermore, it is a general finding\cite{kreibich} that the KLI approximation does not affect the potential asymptotics to leading order. In the actual calculations presented in the following we therefore always use the KLI approximation.

In the OEP (and KLI) scheme the chain rule for functional derivatives \cite{Grabo1997} relates the derivative with respect to the density to the derivatives with respect to the orbitals, 
\begin{equation}
 u_{i\s}\rr= \frac{1}{\varphi^*_{i\s}\rr}\frac{\delta
E_\mathrm{xc}[\{\varphi\}]}{\delta\varphi_{i\s}\rr}.
\end{equation}
From the structure of the OEP equation it further follows that to first order 
\begin{equation}
 \lim_{|\rb|\rightarrow\infty}v_{\mathrm{xc}\s}\rr = \lim_{|\rb|\rightarrow\infty}u_{N_\s \s}\rr, \label{eq.upot_asy}                                                                     
\end{equation}
 i.e., the functional derivative with respect to the HOMO in general determines the potential asymptotics~\cite{kreibich}. Therefore, investigating the HOMO functional derivative
is the key to determining the asymptotic behavior of an orbital dependent functional's xc potential.
When one takes the functional derivative (with respect to the orbital) of a local hybrid one obtains three terms, corresponding to the three addends in Eq.~(\ref{eq.lh_spec}):
\begin{equation}
 u^\mathrm{lh}_{i\s}\rr = u^\mathrm{exx}_{i\s}\rr + u^\mathrm{c-nl}_{i\s}\rr
+u^\mathrm{c-sl}_{i\s}\rr \label{eq.u_allparts}
\end{equation}
Evaluating the asymptotical behavior of each of these three terms for the highest occupied orbital allows one to predict the potential asymptotics.

The first term can be derived directly from Eq.~(\ref{eq.exx}) and reads
\begin{equation}
u^\mathrm{exx}_{i\s}\rr = -\frac{1}{\varphi^*_{i\s}\rr}
\sum_{j=1}^{N_{\s}}\varphi_{j\s}^*\rr\int
\frac{\varphi^*_{i\s}\rrp\varphi_{j\s}\rrp}{|\mathbf{r}-\mathbf{r}'|}\,\mathrm d^3r'
\label{eq.u_exx}
\end{equation}
This term evaluated for the HOMO indeed provides the correct asymptotic behavior~\cite{Grabo1997}
\begin{equation}
u^\mathrm{exx}_{N_\s \s}\rr \underset{|\rb|\rightarrow \infty}{\longrightarrow} 
-\frac{1}{|\rb|}.
\label{eq.u_exx_asy}
\end{equation}
The third term  $u^\mathrm{c-sl}_{i\s}\rr$, on the other hand, does not contribute to the asymptotics of Eq.~(\ref{eq.u_exx_asy}) as it decays exponentially due to its purely (semi-)local nature. 

Evaluating the second term on the right-hand side of Eq.~(\ref{eq.u_allparts}) requires careful consideration. Intuitively, one might
expect that an asymptotically vanishing LMF will surpress any asymptotic contribution of this term to the potential. In the following we check this expectation. Details of the underlying calculation for both LMFs used in this work, i.e., $f_t\rr$ and $f_0\rr$, can be found in Ref.~\cite{iso} and in Appendix~\ref{sec.appendix_details}, Eq.~(\ref{eq.uiso_f0c_omplete}), respectively.

By defining $P\rr= n\rr\,e_\mathrm{x}^\mathrm{sl}\rr$ and $Q\rr= n\rr\,e_\mathrm{x}^\mathrm{ex}\rr$ one can write
\begin{eqnarray}
 u^\mathrm{c-nl}_{i\s}\rr&=&\frac{1}{\varphi^*_{i\s}\rr}\frac{\delta}{\delta \varphi_{i\s}\rr} \int
f\rrp\,\lp(P\rrp -Q\rrp\rp)\,\mathrm d^3r' \notag \\
&=& \frac{1}{\varphi^*_{i\s}\rr}\lp[\int \left(\frac{\delta f\rrp}{\delta\varphi_{i\s}\rr}\right)P\rrp\,\mathrm d^3r'\right.\notag \\
& & +  \int f\rrp\left(\frac{\delta P\rrp}{\delta\varphi_{i\s}\rr
}\right)\mathrm d^3r' \notag \\
& &-\int \left(\frac{\delta f\rrp}{\delta\varphi_{i\s}\rr
}\right)Q\rrp\,\mathrm d^3r'\notag \\
& & \left.-  \int f\rrp\left(\frac{\delta Q\rrp}{\delta\varphi_{i\s}\rr
}\right)\mathrm d^3r'\right]. 
\label{eq.u_nl_prod}
\end{eqnarray}
The first two terms consist of (semi-)local components and thus vanish exponentially. 
Evaluating the third term on the other hand is not as trivial as it contains the nonlocal quantity $Q\rr$ as well the functional derivative of the LMF with respect to the corresponding Kohn-Sham orbital. For the LMFs addressed in this perspective we find that this term does not contribute to the asymptotics either (see Ref.~\cite{iso} and Appendix~\ref{sec.appendix_details} for details). 
Thus, only the fourth term in Eq.~(\ref{eq.u_nl_prod}) is relevant
 in the asymptotic limit,

Thus, only the fourth term in Eq.~(\ref{eq.u_nl_prod}) is relevant
 in the asymptotic limit and therefore 
\begin{eqnarray}
 u^\mathrm{c-nl}_{i\s}\rr&\rightarrow& \frac{1}{2\varphi^*_{i\s}\rr}\lp[f\rr
\sum_{j=1}^{N_{\s}} \varphi^*_{j\s}\rr\notag \int
\frac{\varphi_{i\s}^*\rrp\varphi_{j\s}\rrp}{|\mathbf{
r }-\mathbf{r}'|} \,\mathrm d^3r'\right.\notag \\
& & \left.+\sum_{j=1}^{N_{\s}} \varphi^*_{j\s}\rr \int
f\rrp\frac{\varphi_{i\s}^*\rrp\varphi_{j\s}\rrp}{|\mathbf{r}-\mathbf{r}'|} \,\mathrm d^3r'\rp].
\label{eq.u_nl_prod2}
\end{eqnarray} 
The first term in this equation equals $-u^\mathrm{exx}_{i\s}\rr$ of Eq.~(\ref{eq.u_exx}), locally multiplied by
$f\rr/2$. Due to Eq.~(\ref{eq.f-asy}) it vanishes faster than the leading term of $u_{i \sigma}^\mathrm{lh}$, which is given in Eq.~(\ref{eq.u_exx_asy}). 

The second term, however, is of a different structure, as it evaluates the LMF under the integral. By considering the HOMO level, its asymptotic limit is  
\begin{equation}
u^\mathrm{c-nl}_{N_\s \s}\rr \underset{|\rb|\rightarrow \infty}{\longrightarrow}\half \int
f\rrp\frac{\varphi^*_{N_\s \s}\rrp\varphi_{N_\s \s}\rrp}{|\mathbf{r}-\mathbf{r}'|}\,\mathrm
d^3r' . \label{eq.u_nl_asy}
\end{equation}
This corresponds to a Hartree-like potential caused by the spin-orbital density of the HOMO averaged over all space, with the LMF as a weighting function. Thus, this term  
gives a finite contribution in the asymptotic limit despite of Eq.~(\ref{eq.f-asy}).

Now, when adding the asymptotically significant components, 
Eq.~(\ref{eq.u_exx_asy}) and Eq.~(\ref{eq.u_nl_asy}),
for the evaluation of Eq.~(\ref{eq.upot_asy}), we arrive at
\begin{equation}
v_{\mathrm{xc} \s}\rr \underset{|\rb|\rightarrow
\infty}{\longrightarrow}-\frac{\gamma_\s}{|\rb|}. \label{eq.pot_compl_asy} 
\end{equation}
Here, the parameter $\gamma_\s$ denotes the reduced slope of the potential asymptotics, which can numerically be extracted from a self-consistent Kohn-Sham calculcation via
\begin{equation}
\gamma_\s = 1-\half\int f\rr|\varphi_{N_\s \s}\rr|^2\,\mathrm d^3r. \label{eq.gamma}
\end{equation}
Eq.~(\ref{eq.gamma}) is a central result of this work, as it demonstrates that a local hybrid of the form of Eq.~(\ref{eq.lh_spec}) does not lead to the exact asymptotic behavior of the xc potential.
Eq.~(\ref{eq.pot_compl_asy}) holds 
for all $f\rr$ that vanish in the asymptotic limit and for which the third term of Eq.~(\ref{eq.u_nl_prod}) does not contribute to the asymptotics of the functional derivative $u^\mathrm{c-nl}_{i\s}\rr$, i.e., under very general conditions.
Further details of the calculation, specifically  regarding the question of the xc potential asymptotics in different spin-channels, are given in Appendix~\ref{sec.appendix_details}. 

The LMF is limited between  $0 \le f\rr \le 1$ and therefore the asymptote is bound between $ \half  < \gamma_\s \le 1$. Consequently, the exact value
 $\gamma_\s = 1$ can only be reached by setting $f\rr=0 \, \forall \, \rb$, which corresponds to the trivial case of using EXX, ``as is'' or combined with a purely (semi-)local correlation functional. 

A different extreme case, $f\rr=1 \, \forall \, \rb$, does not, as one could na\"{\i}vely believe due to Eq.~(\ref{eq.gamma}), lead to $\gamma_\s=\half$. Here, we have to take the neglected
 first term of Eq.~(\ref{eq.u_nl_prod2}) into account again, and from this we see that $\gamma_\s$ actually vanishes. This is to be expected, since in this case the local hybrid reduces to a purely
(semi-)local functional. 

\begin{figure}[t]
  \includegraphics[width=8cm,trim=0mm 20mm 0mm 150mm]{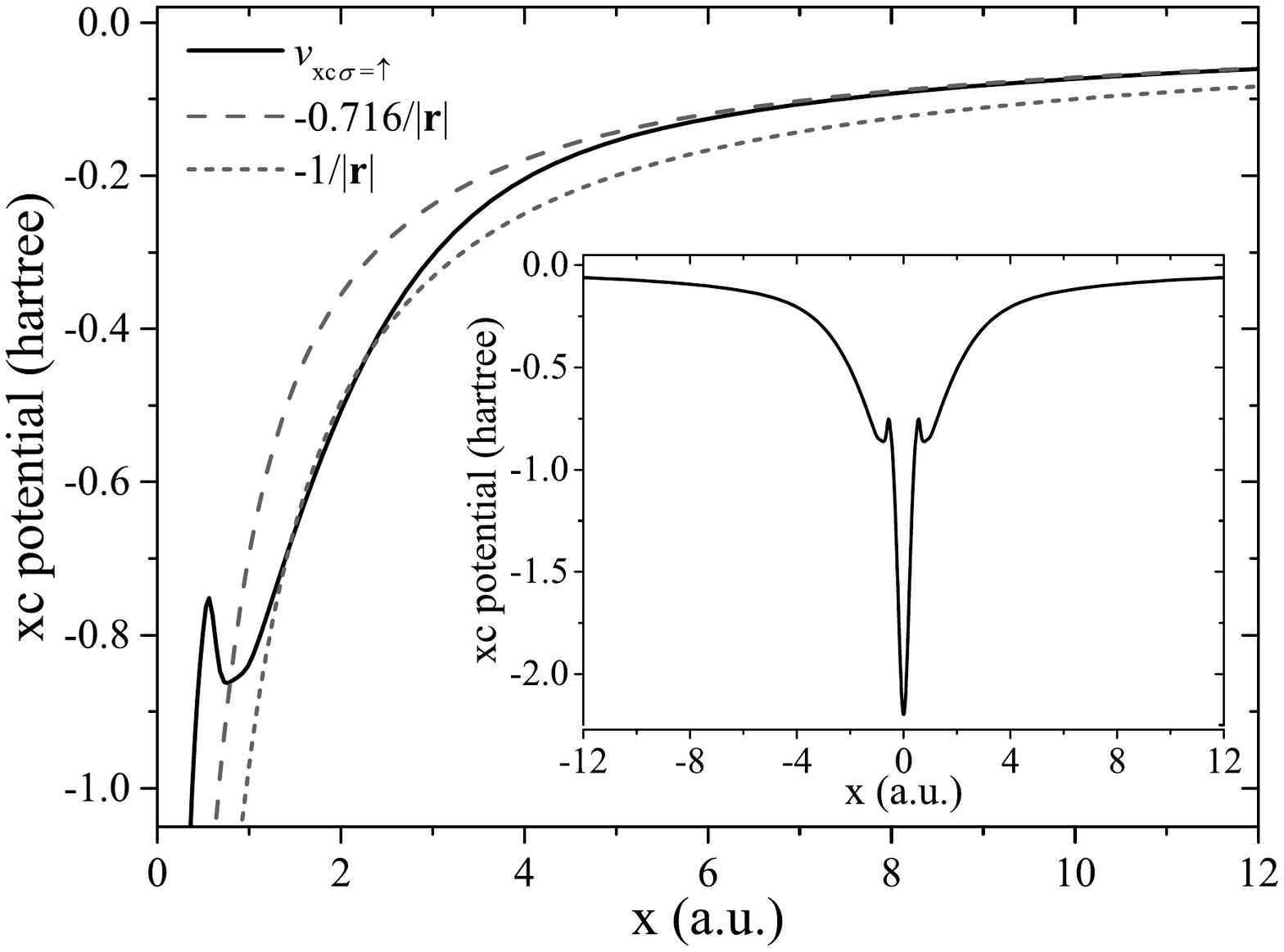}\\
  \caption{The xc potential $v_{\mathrm{xc}\up}\rr$ for the C atom along the $x$-axis, 
(see Appendix~\ref{sec.appendix_darsec} for definition), computed using $f_t\rr$ with $c=0.5$. Also displayed is the asymptotic curve according to Eq.~(\ref{eq.pot_compl_asy}) with $\gamma_\up(c=0.5)=0.716$ and the correct asymptotic $-1/r$. The inset shows the potential plotted along the complete axis.}
\label{fig.Cupisox}
\end{figure}
Fig.~\ref{fig.Cupisox} shows a numerical verification of the above analytical considerations (see Appendix \ref{sec.appendix_darsec} for numerical details). It depicts the xc (KLI) potential corresponding to the local hybrid of Eq.~(\ref{eq.fiso}) in comparison with the asymptotic decay according to Eq.~(\ref{eq.pot_compl_asy}) and Eq.~(\ref{eq.gamma}) for the carbon atom. An additonal curve indicates the exact $-1/r$ decay, which is clearly not reached. The xc potential, instead of decaying with $\gamma_\up=1$ as one would intuitively expect~\cite{Arbuznikov2007}, approaches the predicition of Eq.~(\ref{eq.gamma}) ($\gamma_\up=0.716$) quite rapidly.

\begin{figure}[t]
  \includegraphics[width=8cm,trim=0mm 20mm 0mm 150mm]{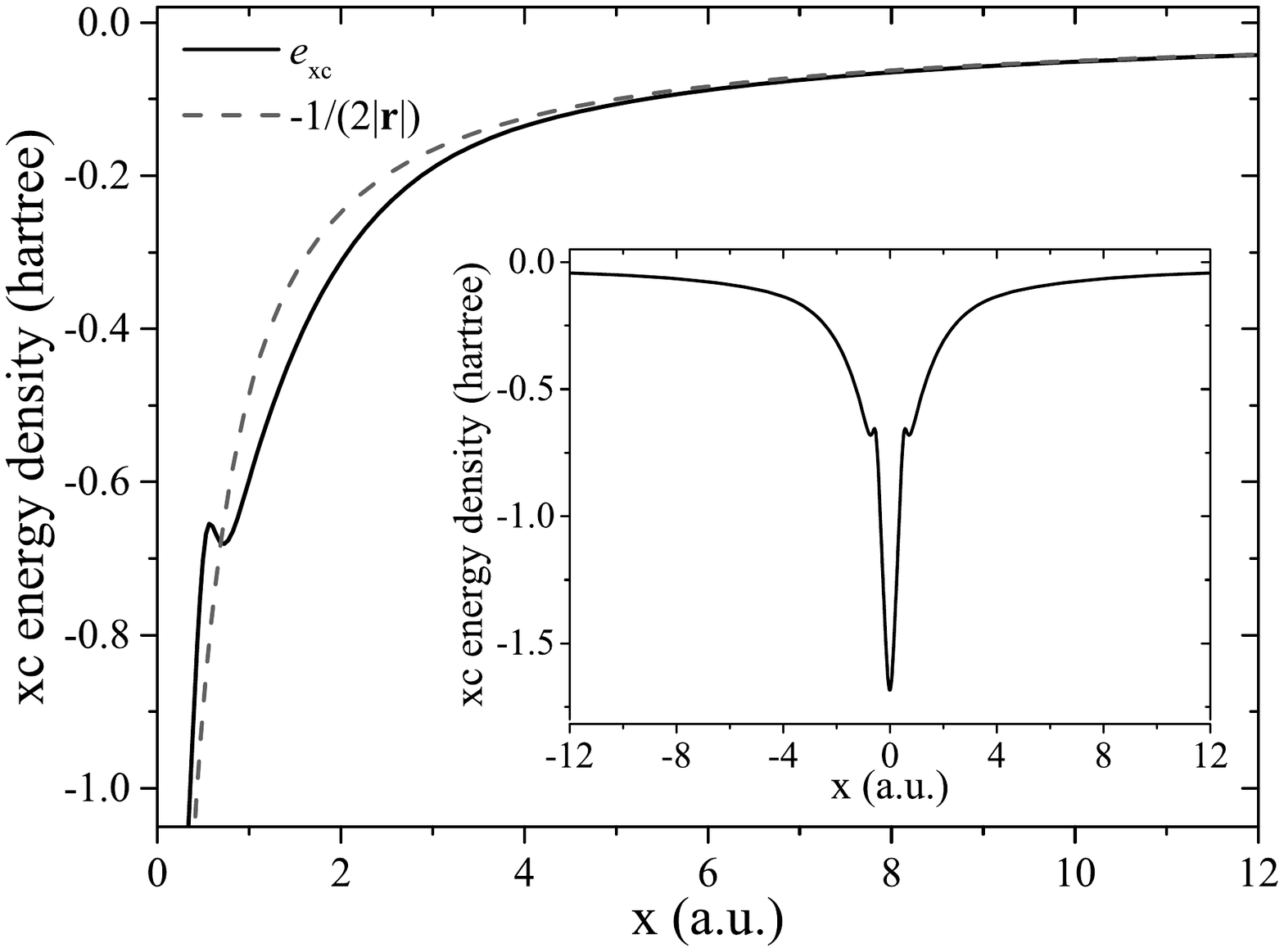}\\
  \caption{xc energy density per particle $e_\mathrm{xc}\rr$ for the C atom along the $x$-axis (see Appendix~\ref{sec.appendix_darsec} for definition), computed using $f_t\rr$ with $c=0.5$. Also displayed is the corresponding
asymptotic slope of $-1/(2r)$. The inset shows the energy density plotted along the complete axis.}
\label{fig.Cexcx}
\end{figure}
Fig.~\ref{fig.Cexcx} shows the xc energy density $e_\mathrm{xc}\rr$ for the same system in comparison to its correct asymptotic of $-1/(2r)$. Clearly the xc energy density shows the correct asymptotic, cf.\ Eq.~(\ref{eq.exc-asy}). 
We thus see that while the behavior of $e_\mathrm{xc}$ can directly be controlled via the LMF in Eq.~(\ref{eq.lh_spec}), the process of finding the local xc potential via functional differentiation leads to non-local evaluations of the LMF that decisively impact the potential's asymptotics.

A physically meaningful quantity closely related to the asymptotics of the xc potential is the highest occupied eigenvalue $\eps_\mathrm{ho}$. Table~\ref{tbl.IP} shows $-\eps_\mathrm{ho}$ compared to the experimental IP for the carbon atom for different functionals, together with the corresponding value of $\gamma_\s$ of the xc potential from Eq.~(\ref{eq.gamma}). 

\begin{table}[h]
\small
  \caption{ Comparison of the highest occupied Kohn-Sham eigenvalue $-\eps_\mathrm{ho}$ to the experimental vertical IP~\cite{HandChemPhys92} for the C ($\eps_{4\up}$) and F atoms ($\eps_{4\dn}$) using different functionals. All values are in hartree. 
}
  \label{tbl.IP}
  \begin{tabular*}{0.5\textwidth}{@{\extracolsep{\fill}}lllll}
    \hline
    System & functional & $\gamma_{\s_\mathrm{ho}}$ & $-\eps_\mathrm{ho}$ & exp. IP \\
    \hline
    C 	& LSDA	 		& -- 		&0.2249 &    0.4138\\
	& $f_t\rr(c=0)$ 	& 0.6098	&0.2740 &	 \\
	& $f_t\rr(c=0.5)$	& 0.7162	&0.3067 &	 \\
	& $f_t\rr(c=1.0)$	& 0.7678	&0.3302 &	 \\
	& $f_t\rr(c=2.5)$	& 0.8441	&0.3688 &	 \\
	& $f_t\rr(c=5.0)$	& 0.8966	&0.3970 &	 \\
	& $f_0\rr$		& 0.8309	&0.3530 &	 \\
	& EXX			& 1.0000	&0.4378 &	 \\
    \hline
    F 	& LSDA	 		& -- 		&0.3808 &    0.6403\\
	& $f_t\rr(c=0)$ 	& 0.5055	&0.3810 &	 \\
	& $f_t\rr(c=0.5)$	& 0.6665	&0.4724 &	 \\
	& $f_t\rr(c=1.0)$	& 0.7390	&0.5269 &	 \\
	& $f_t\rr(c=2.5)$	& 0.8365	&0.6060 &	 \\
	& $f_t\rr(c=5.0)$	& 0.8971	&0.6570 &	 \\
	& $f_0\rr$		& 0.7927	&0.5798 &	 \\
	& EXX			& 1.0000	&0.6779 &	 \\
    \hline
  \end{tabular*}
\end{table}

\begin{figure}[h]
  \includegraphics[width=8cm,trim=0mm 20mm 0mm 150mm]{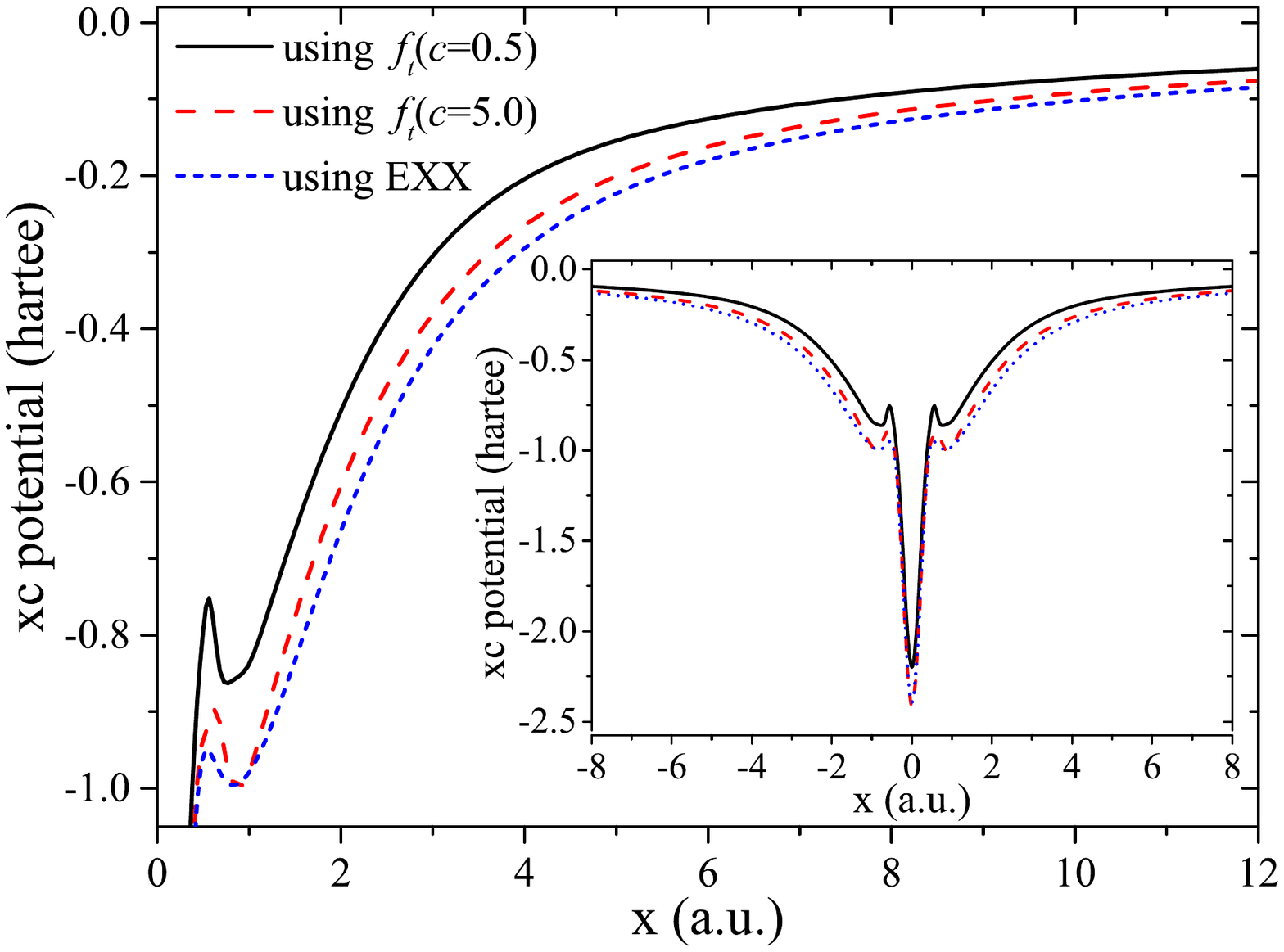}\\
    \caption{Asymptotics of the xc potential $v_{\mathrm{xc}\up}\rr$ for the C atom along the $x$-axis, computed with pure EXX and local hybrids using $f_t\rr$ from Eq.~(\ref{eq.fiso}) with parameters c=0.5 and c=5.0. Also displayed are the corresponding asymptotics for $\gamma_\up(EXX)=1$,  $\gamma_\up(c=0.5)=0.716$ and  $\gamma_\up(c=5.0)=0.897$, and the complete potential in the inset.} 
\label{fig.C_exx_x}
\end{figure}

The LSDA, as generally known, significantly underestimates the IP due to the wrong potential asymptotics and the 
inherent self-interaction-error. Using pure EXX with the correct asymptotic decay and no self-interaction-error leads to a much better prediction of the IP.
When employing a local hybrid with the LMF $f_t\rr$, the explicit dependence on the parameter $c$ becomes evident: With growing $c$, the asymptotic value $\gamma_\s$ grows and the description of the IP improves.
Fig.~\ref{fig.C_exx_x} sheds further light on the situation. It shows potentials of local hybrids which are all based on Eq.~(\ref{eq.fiso}) but use different values of $c$. Growing values of $c$ increase the amount of EXX and lead to an overall deeper potential. This explains that the eigenvalues become more negative.

We thus see that while all of the local hybrids used here can (so far, see caveat in the next section) be thought of as being one-electron self-interaction-free, they show different potential asymptotics and their highest occupied eigenvalues predict the IP with significantly different reliability. The relation between freedom from self-interaction, potential asymptotics, and physical interpretability of the highest occupied eigenvalue as the negative IP is therefore much less clear than intuitively believed. This observation also calls for taking a closer look at the iso-orbital indicator $g\rr$ that is used in enforcing freedom from self-interaction. This is the topic of the next section.

\section{The implications of orbital nodal planes}\label{sec.theory_np}

As explained in the preceding 
sections, many local hybrids and other functionals such as meta-GGAs rely on the function $g\rr$ tending to $1$ to detect regions of space in which a single orbital shape dominates the density, and then, e.g., correct for self-interaction in such regions. However, a first caveat that one has to take note of is that $g\rr \rightarrow 1$ holds for one-particle densities of ground-state character. This is a possibly 
far reaching restriction for the use of $g\rr$ because electron orbital densities typically have nodes, i.e., are {\it not} of ground-state character.

As a particular example, consider a case where the HOMO is a $p$-orbital with an azimuthal quantum number $m$ and is expressed in spherical coordinates as $\varphi_\mathrm{ho}\rr = R(r,\theta)e^{i m \phi}$. At the region where the density is dominated by the HOMO, $\tau_\mathrm{W}\rr = \frac{|\nabla n\rr|^2}{8 n\rr} = \frac{1}{2} |\nabla \sqrt{n\rr}|^2 \rightarrow \frac{1}{2} | \nabla R(r,\theta) |^2$. However, $\tau\rr \rightarrow \frac{1}{2} |\nabla \varphi_\mathrm{ho}\rr|^2 = \frac{1}{2} ( | \nabla R(r,\theta) |^2 + m^2 R^2(r,\theta) / (r \sin \theta)^2)$. As a result, while for $m=0$, $\tau\rr$ identically equals $\tau_\mathrm{W}\rr$, for $m = \pm 1$ this is no longer the case, but $\tau\rr$ and $\tau_\mathrm{W}$ approach each other asymptotically, as the $m^2$-dependent term of $\tau\rr$ decays to zero.

One may counter-argue that this restriction is not so severe because in density functional construction the condition $d\rr \rightarrow 1$ is mostly used to detect those regions of space in a finite system which 
are far from all nuclei, where the density decays nodelessly, and there $\tau\rr \rarr \tau_\mathrm{W}\rr$ even in the example above. 

However, we show below that even in such regions the condition $\tau\rr \rarr \tau_\mathrm{W}\rr$ can be violated. 
This leads to a second caveat about the reliability of the $g\rr$ indicator. It is rooted in the existence of orbital densities that have nodal planes or nodal axes. Fig.~\ref{fig.twt} illustrates the case. 
It shows $g\rr$ evaluated in the $(xz)$-plane for the carbon atom (see Appendix \ref{sec.appendix_darsec} for numerical details, including grid setup). The density here was obtained using $f_t\rr(c=0.5)$, but the density features relevant here are not sensitive to functional details. The important observation is that $g\rr$ approaches $1$ in the asymptotic limit in every direction -- except for in the vicinity of $x=0$. 
\begin{figure}[t]
  \includegraphics[width=8cm,trim=0mm 20mm 0mm 150mm]{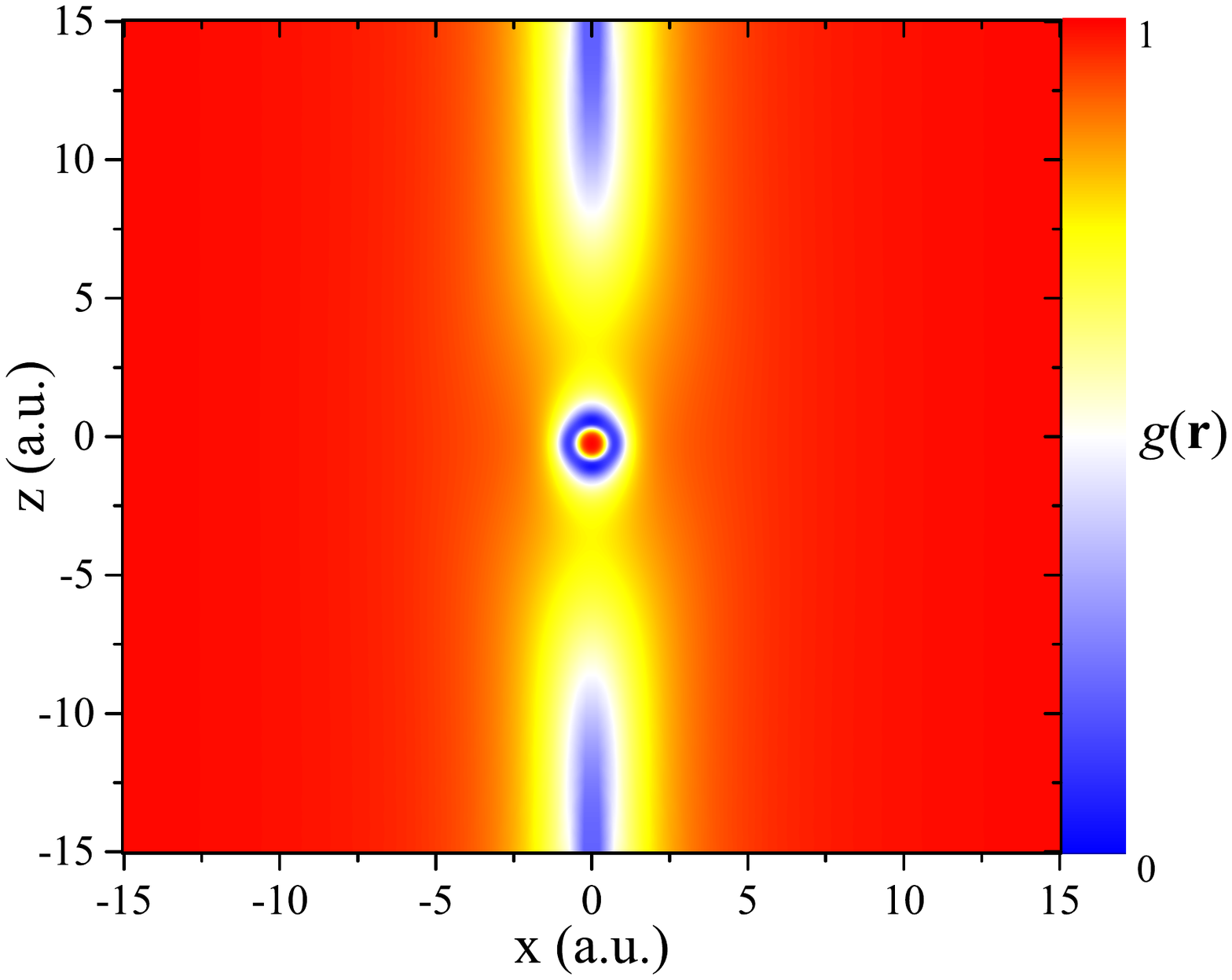}\\
    \caption{The function $g\rr=\frac{\tau_W\rr}{\tau\rr}$ on the numerical grid for the C atom 
     }\label{fig.twt}
\end{figure}

The first step towards an understanding of this finding is to note that the $z$-axis is a nodal axis, being the intersection of the nodal planes of the two HOMOs of carbon, which are degenerate and of $p$-orbital character. 

The consequences of the existence of nodal planes can be studied analytically. To this end we look at a schematic density that is dominated by the HOMO $\varphi_\mathrm{ho}\rr$, but also take the next lower lying orbital $\varphi_\mathrm{ho-1}\rr$ into account, i.e. $n\rr \sim |\varphi_\mathrm{ho}\rr|^2+|\varphi_\mathrm{ho-1}\rr|^2$.
With this ansatz one finds
\begin{equation}
 \tau_\mathrm{W} \sim \frac{\lp(\nabla |\varphi_\mathrm{ho}|^2 + \nabla |\varphi_\mathrm{ho-1}|^2 \rp)^2}{8\lp(|\varphi_\mathrm{ho}|^2+|\varphi_\mathrm{ho-1}|^2\rp)}
\end{equation}
and
\begin{equation}
 \tau \sim \half{\lp(|\nabla \varphi_\mathrm{ho}|^2 + |\nabla \varphi_\mathrm{ho-1}|^2 \rp)}.
\end{equation}
These two terms combined and evaluated on or close to a nodal plane (denoted by $\underset{n.p.}{\longrightarrow}$), 
where $\varphi_\mathrm{ho}\rarr0$, yield
\begin{equation}\label{eq.twt.np}
 \frac{\tau_\mathrm{W}}{\tau} \underset{n.p.}{\longrightarrow} \frac{|\nabla \varphi_\mathrm{ho-1}|^2}{|\nabla \varphi_\mathrm{ho}|^2+|\nabla \varphi_\mathrm{ho-1}|^2} < 1.
\end{equation}
Even though $\varphi_\mathrm{ho}$ vanishes on the nodal plane, its gradient still yields a finite value and keeps the function $g\rr$ from approaching 1. 

Fig.~\ref{fig.twt} shows that the deviation from $1$ has a noticeable spatial extension of a few a.u. This raises the question of how well the use of the iso-orbital indicator $g\rr$ leads to freedom from self-interaction, as in some regions that so far have been considered as iso-orbital  ones, e.g., all space far from the system's center, self-interaction effects may not be eliminated fully when the indicator aberrates due to the presence of a nodal plane or axis. 
A different interpretation of Fig.~\ref{fig.twt} would be to reconsider one's expectation of where iso-orbital regions are, or what they are. The traditional point of view has been that all space far from a finite system's center is of iso-orbital nature. Fig.~\ref{fig.twt} and Eq.~(\ref{eq.twt.np}) may be interpreted to show that this is not the case when the HOMO has a nodal plane/axis extending to infinity. From this perspective one might say that $g\rr$ does exactly what it is supposed to be doing, i.e., it indicates that the nodal plane region is not of iso-orbital character. Yet, also from this perspective Fig.~\ref{fig.twt} reveals a surprising finding, namely that even infinitely far from a finite system's center, the density may not be of iso-orbital character.

The nodal plane observation also forces us to take a yet closer look at the central topic of this perspective, the potential asymptotics.
Nodal planes can influence the asymptotics of a local hybrid's xc potential in two ways. 
First, it has been argued that all orbital-dependent functionals show non-vanishing asymptotic constants in their xc potential along nodal planes of the highest occupied Kohn-Sham orbital that extend to infinity. This was first discussed in Refs.~\cite{goerlingasymp,KP03a} for the case of pure EXX, and the occurring shift was determined to be 
\begin{equation}
 C_\s = \bar{v}_{\mathrm{xc} M_\s \s}-\bar{u}_{\mathrm{xc} M_\s \s} \label{eq.nonvanishingc},	
\end{equation}
with $\bar{v}_{\mathrm{xc} i \s}= \int \varphi_{i\s}^*\rr v_{\mathrm{xc} \s}\rr\varphi_{i\s}\rr\,\mathrm d^3r$ and  $\bar{u}_{\mathrm{xc} i \s}= \int\varphi_{i\s}^*\rr u_{\mathrm{xc} i\s}\rr\varphi_{i\s}\rr\,\mathrm d^3r$. 
The index $M_\s$ denotes the highest lying Kohn-Sham orbital that does not show a vanishing 
spin-orbital density along the nodal plane of the HOMO. Since Eq.~(\ref{eq.nonvanishingc}) follows from 
the KLI (OEP) equation without referring to a specific functional, nonvanishing asymptotic constants on nodal planes of the HOMO are expected on rather general grounds.

\begin{figure}[t]
  \includegraphics[width=8cm,trim=0mm 20mm 0mm 150mm]{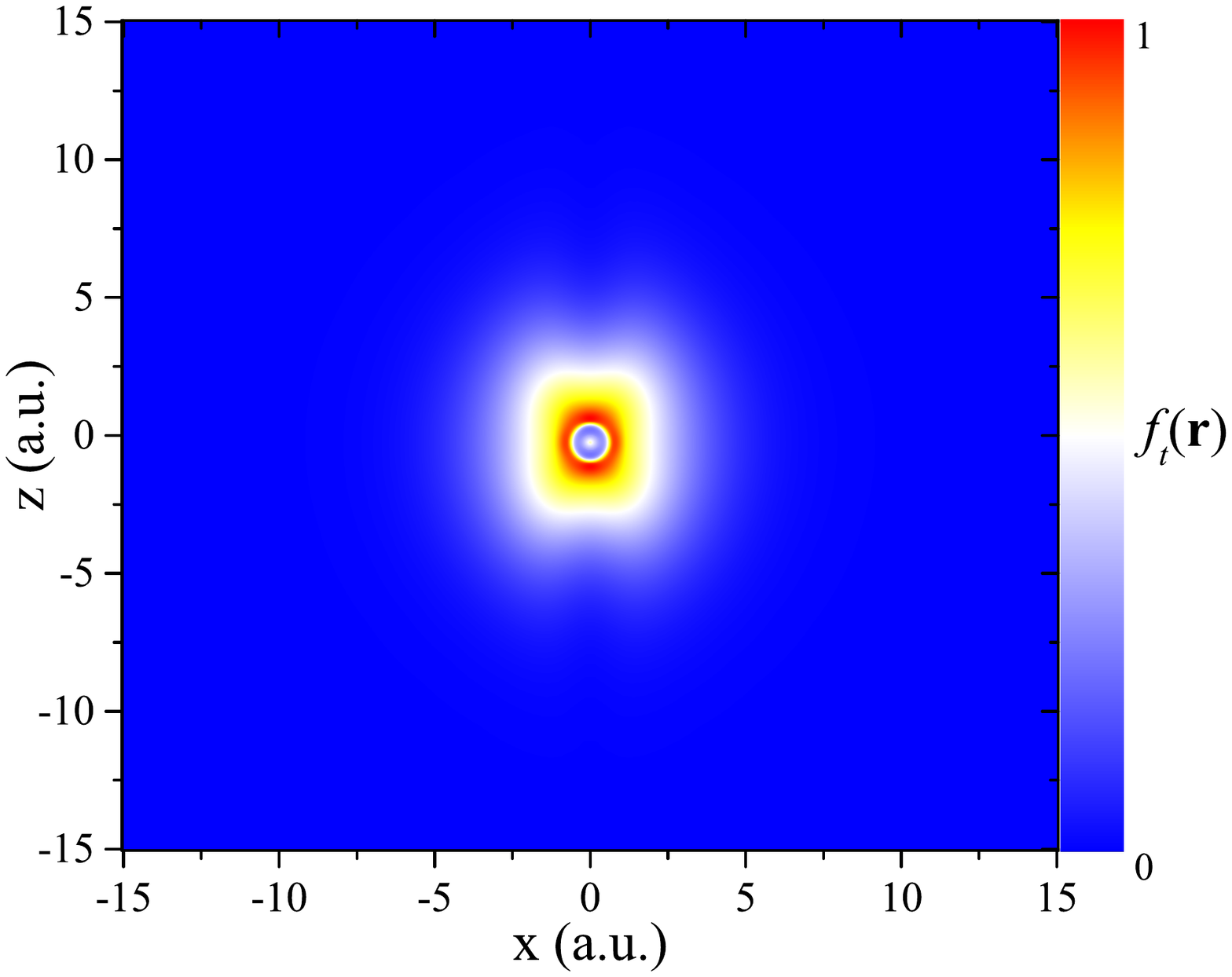}\\
    \caption{The LMF $f_{t}\rr = \frac{1 - \frac{\tau_W\rr}{\tau\rr} \zeta^2\rr}{1+ct^2\rr}$, evaluated with $c=0.5$, on the numerical grid for the C atom.}\label{fig.ft2d}
\end{figure}
\begin{figure}[h!]
  \includegraphics[width=8cm,trim=0mm 0mm 0mm 150mm]{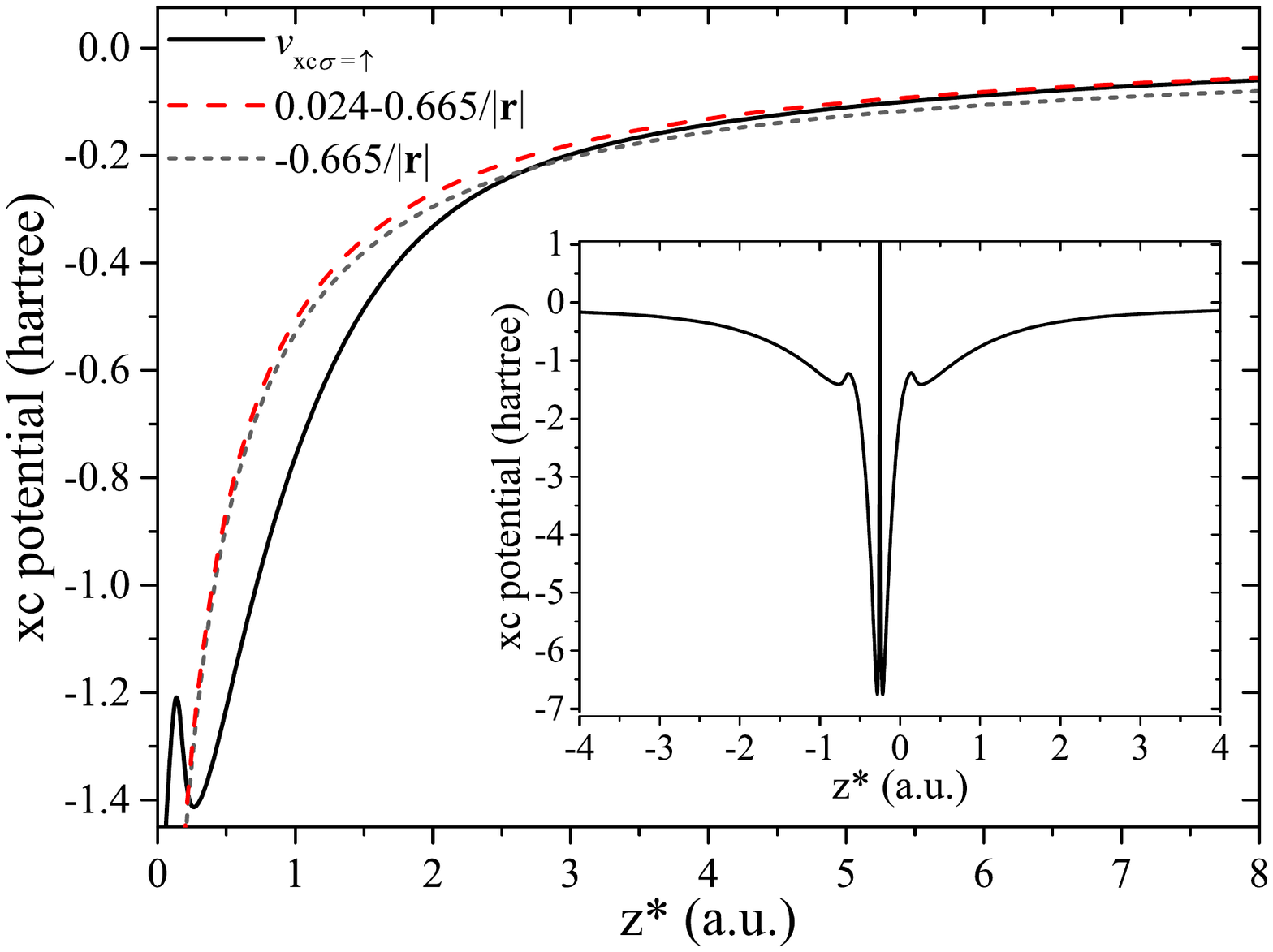}\\
    \caption{Asymptotics of the xc potential $v_{\mathrm{xc}\up}\rr$ for the F atom along the (projected) $z$-axis (denoted $z^*$, see Appendix~\ref{sec.appendix_darsec} for definition), computed using $f_t\rr$ with the parameter c=0.5. }
\label{fig.F_iso_z}
\end{figure}

Second, the fact that $g\rr \rarr 1$ is not guaranteed on a nodal plane can also affect the potential. For the sake of clarity, we again discuss this effect for the specific example of local hybrids. When the LMF tends to zero on the nodal plane, i.e., $f\rr \underset{n.p.}{\rightarrow} 0$ and Eq.~(\ref{eq.f-asy}) is obeyed, then the non-vanishing constant of Eq.\ (\ref{eq.nonvanishingc}) is the only effect. An example for this case is the LMF 
$f_t\rr$ with a finite value of the parameter $c$. It is depicted in Fig.~\ref{fig.ft2d} for the C atom density in the $(xz)$-plane, and one sees that there are no asymptotic features. 
This is because the reduced density gradient in the denominator causes $f_t\rr$ to vanish in the asymptotic limit, regardless of the occurrence of a nodal plane. The potential decays like $-\gamma_\s/r$ in all directions, but along the $z$-axis a nonvanishing constant 
\begin{equation}
v_{\mathrm{xc} \s}\rr \underset{n.p.}{\longrightarrow}C_\s-\frac{\gamma_\s}{|\rb|} \label{eq.potnonvanishing} 
\end{equation}
appears. This is shown in Fig.~\ref{fig.F_iso_z} for $f_t\rr(c=0.5)$ and the F atom. One can clearly see how $v_{\mathrm{xc}\up}\rr$ decays with $\gamma_\up=0.6650$, but,
instead of zero, approaches a constant of $C_\up=0.0244$, in agreement with Eq.~(\ref{eq.nonvanishingc}).

\begin{figure}[h!]
  \includegraphics[width=8cm,trim=0mm 0mm 0mm 150mm]{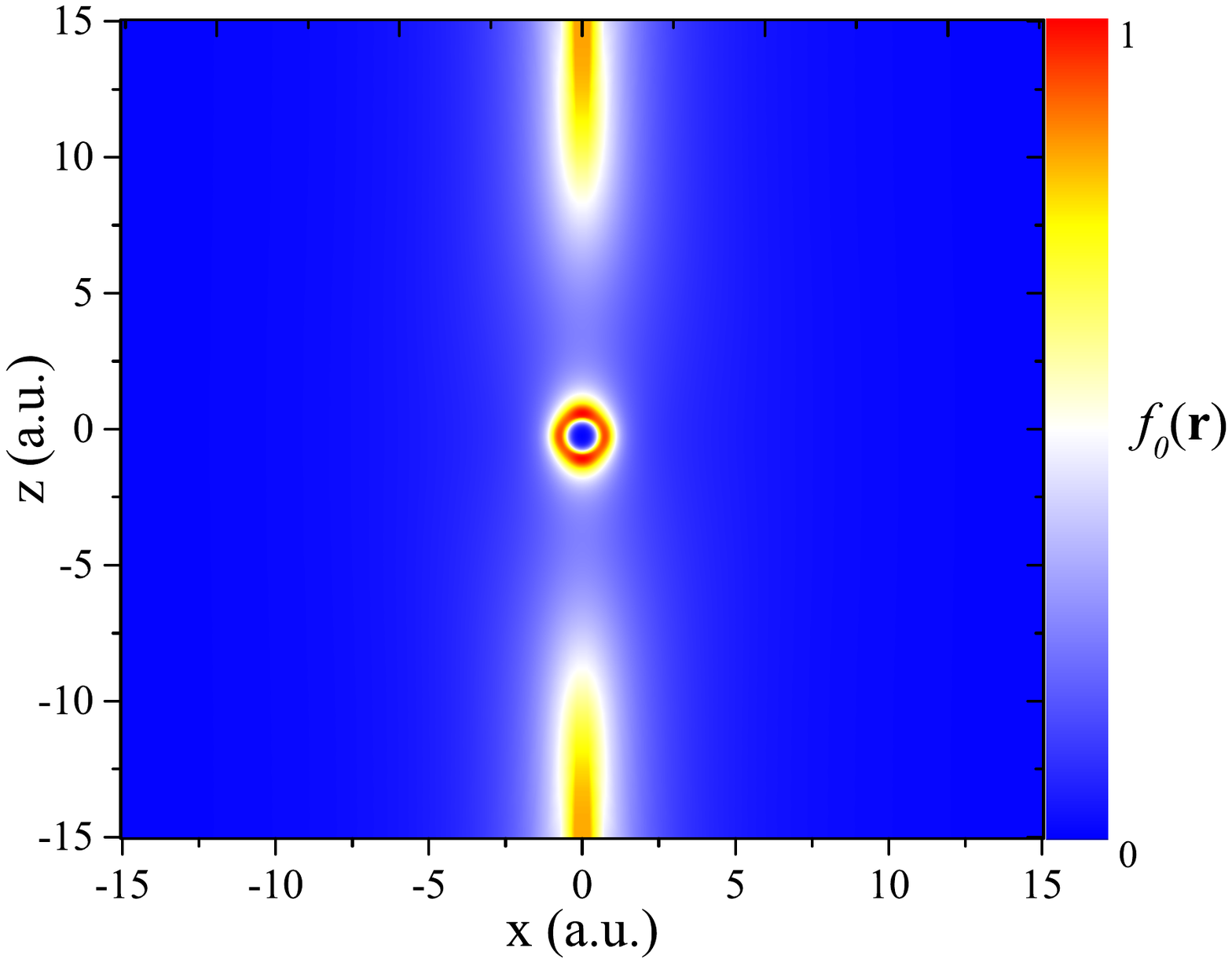}\\
    \caption{The LMF $f_{0}\rr =1- \frac{\tau_W\rr}{\tau\rr}$ on the numerical grid for the C atom.}\label{fig.f02d}
\end{figure}

A different situation occurs when $f\rr \underset{n.p.}{\nrightarrow} 0$, i.e., the behavior of the indicator function along a nodal plane/axis 
of the HOMO prevents the LMF from reaching 
its intended limit. This happens, e.g., for $f_0\rr$ or $f_t\rr(c=0)$ and is depicted in  Fig.~\ref{fig.f02d}, again for the C atom. The occurrence of a nodal axis 
here very clearly affects the LMF. Since in this case Eq.~(\ref{eq.f-asy}) is violated in the direction of the $z$-axis, the previous derivations cannot be used to predict the potential's asymptotic behavior. 
However, we have numerically checked the xc potential's behavior. On the nodal axis it neither tends to $-1/r$, nor to $C_\s-\frac{\gamma_\s}{|\rb|}$, but rather tends to some other value. Thus, the nodal axis in this case has a very noticeable influence on the potential asymptotics, which is hard to predict a priori.

\section{Conclusions}\label{sec.conclusions}

With local hybrid functionals serving as an explicit example we have argued that freedom from self-interaction in the sense of Eq.\ (\ref{eq.si-error}) does not necessarily lead to the expected $-1/r$ decay of the local Kohn-Sham xc potential. We have further argued that the ratio of the von Weizs\"acker kinetic energy density to the positive Kohn-Sham kinetic energy density, which is frequently used in functional construction for indicating iso-orbital regions and eliminating self-interaction effects in these, may not serve its intended purpose because it is very sensitive to excited state features such as orbital nodal planes that are present in Kohn-Sham orbitals that construct 
ground-state densities of many-electron systems.

These findings have immediate and somewhat discomforting consequences for the local hybrid approach. For a large class of functionals one has to accept that the correct long-range xc potential simply cannot be obtained. This observation plays a role in explaining why it is very hard to construct a local hybrid that yields good binding energetics and physically meaningful eigenvalues with the same functional form and set of parameters\cite{iso}. However, the impaired relation between self-interaction and the xc potential's asymptotics, and also the impact of nodal planes, stand in a context that is much larger than the local hybrid one. The iso-orbital indicator $g\rr$ has been used in many functionals, not only local hybrids. Nodal planes are known to impact the exact exchange potential in surprising ways\cite{goerlingasymp,KP03a}. 
They have appeared here as a prominent feature in kinetic energy ratios, and we expect\cite{thilo} that they are playing a much larger role in the exchange potential than has been realized so far. The observation that a one-electron self-interaction-free energy can go together with a potential that does not fall of like $-1/r$ is not only a feature of local hybrids, but has also been reported for a ``scaled down'' version of the Perdew-Zunger self-interaction correction \cite{scaleddown}. One may therefore wonder whether semi-local indicator functionals are in some sense incompatible with the fully non-local self-interaction correction that is achieved by EXX or full Perdew-Zunger-type correction approaches. It has also been pointed out recently\cite{tdsicjcp} that Eq.\ (\ref{eq.si-error}) itself, which is the basis of the present definition of one-electron self-interaction, leads to questions when evaluated for orbital densities, because $E_\mathrm{xc}[n]$ is intended to be used with ground state densities, whereas orbital densities are excited state densities. Further conceptual questions about Eq.~(\ref{eq.si-error}) relate to its inherent identification of orbitals with electrons and its unitary variance \cite{PZ'81,GOEP,siccomplex}. The success of self-interaction corrections schemes that rely on Eq.~(\ref{eq.si-error}) 
tells us that the equation is meaningful. However, the sum of the insights into its limitations that emerged over the years suggests that there is more to the question of self-interaction in density functional theory. 

While the above considerations point out areas that require further thought and work, one should also note that there have been developments in DFT that shine a bright light into the future. The concept of many-electron self-interaction \cite{manyelsiyang,manyelsiperdew} is not as straightforward to use as Eq.\ (\ref{eq.si-error}), but it avoids the conceptual questions that are associated with this equation. 
Range-separated hybrids yield the correct asymptotic potential and have proven to be a very successful concept, without being self-interaction-free~\cite{SavinFlad_IJoQC56_1995, LeiningerStollEtAl_CPL275_1997,camb3lypCT,SongHirosawaEtAl_TJoCP126_2007,livshits07,wb97x,RohrdanzMartinsEtAl_TJoCP130_2009,steinprl,Refaely-AbramsonBaerEtAl_PRB84_2011, KoerzdoerferSearsEtAl_TJoCP135_2011,akjcpcom,RohrdanzHerbert_TJoCP129_2008,lkskbook,tuningperspective,Pandey2012,Refaely-Abramson2012}. There have been successful functional constructions that can be seen as combinations of the local hybrid and the range-separation idea \cite{Janesko2008,haunschild2010}. Ensemble corrections \cite{kraislerprl} allow to extract information from functionals in an unexpected way, and can, e.g., further improve IP prediction.
Finally, it has recently been shown\cite{newgga} that a new type of a generalized gradient approximation can show features that were so far thought of as being associated only with exact exchange, such as step structures and surprising nodal plane features \cite{thilo}, and understanding potentials in terms of xc charges has provided new insights \cite{kohut,Gidopoulos2012,Andrade2011}. Therefore, the battle against DFT's old foe, the self-interaction error, and its surprisingly independent side-kick, the wrong potential fall-off, is far from being lost.

\section*{Acknowledgements}
 We acknowledge financial support from the German-Israeli Science foundation. T.S.\ acknowledges support from the Elite Network of Bavaria (``Macromolecular Science'' program). S.K.\ acknowledges support from Deutsche Forschungsgemeinschaft. E.K.\ is a recipient of the Levzion scholarship. L.K.\ acknowledges support from the Lise Meitner Minerva Center for Computational Chemistry. 

\appendix

\section{Numerical details}\label{sec.appendix_darsec}

We used the all-electron code \darsec~\cite{Makmal2009} for all calculations presented in this perspective.
This code exploits the rotational symmetry of diatomic molecules along the interatomic axis $z$, 
treating the azimuthal  
angle $\phi$ analytically and thus effectively reducing the problem of solving the Kohn-Sham equations to two dimensions. The equations are represented on a real-space grid of prolate-spheroidal coordinates. In such a coordinate system, the nuclear position(s) coincide with the focal point(s) of the grid located at $z= \pm R/2$, with $R$ being the bond length of the diatomic molecule. This is the case also for calculations of single atoms: the position of the nucleus is not equivalent to the origin of the coordinate system, but is located at $z=-R/2$, where $R$ was set to be 0.5 a.u. 
E.g., the C atom in our plots is centered at $\rb_C=(x,z)=(0,-0.25)$.
The $x$-axis is defined as perpendicular to the $z$-axis, crossing the latter at $z=0$, i.e., at a point being equidistant from the focal points of the grid (see Ref.~\cite{Makmal2009} for details). 

In order to avoid numerical instabilities due to singularities in the Laplacian, the grid was chosen such that it does not include the actual $z$-axis, i.e. the interatomic axis. As a consequence, in this direction all quantities can only be plotted along a projected $z*$-axis, which takes into account all grid points that are closest to the actual $z$-axis. Since the discrepancy between the projected and the real $z$-axis decreases with
increasing number of grid points, we made sure that the difference between $z$ and $z*$ is small by choosing sufficiently dense and large grids.

\section{The asymptotic decay of the exchange-correlation potential in detail}\label{sec.appendix_details} 
In the following, we present considerations about the asymptotics of the xc potential in the spin channel that carries the global HOMO ($\s_\mathrm{ho}$), as compared to
the other spin channel ($\bar{\s}_\mathrm{ho}$). Sec.~\ref{sec.theory_asy} used the condition that $f\rr$ needs to vanish at a sufficient rate 
in the derivation of Eq.~(\ref{eq.pot_compl_asy}). In the present work, we investigated two possibilities for the decay of the LMF.

First, $f_t\rr$ for a finite value of the parameter $c$ vanishes exponentially because 
$f_t\rr \sim t^{-2}\rr \sim e^{-\frac{2}{3} \sqrt{-2\eps_\mathrm{ho}}r}$. In this case, all individual terms in each functional derivative, $u^{\mathrm{c-nl}}_{i\s}$ (see Eq.~(\ref{eq.u_nl_prod})), 
vanish exponentially in the asymptotic limit as well, except for the second term in Eq.~(\ref{eq.u_nl_prod2}). Eventually, this remaining term is responsible for the reduced asymptotic decay of Eq.~(\ref{eq.pot_compl_asy}) due to the nonlocal evaluation of $f_t\rr$. 
Consequently, the xc potential in both spin channels decays with $-\gamma_\s/r$. 

\begin{figure}[h!]
  \includegraphics[width=8cm,trim=0mm 0mm 0mm 150mm]{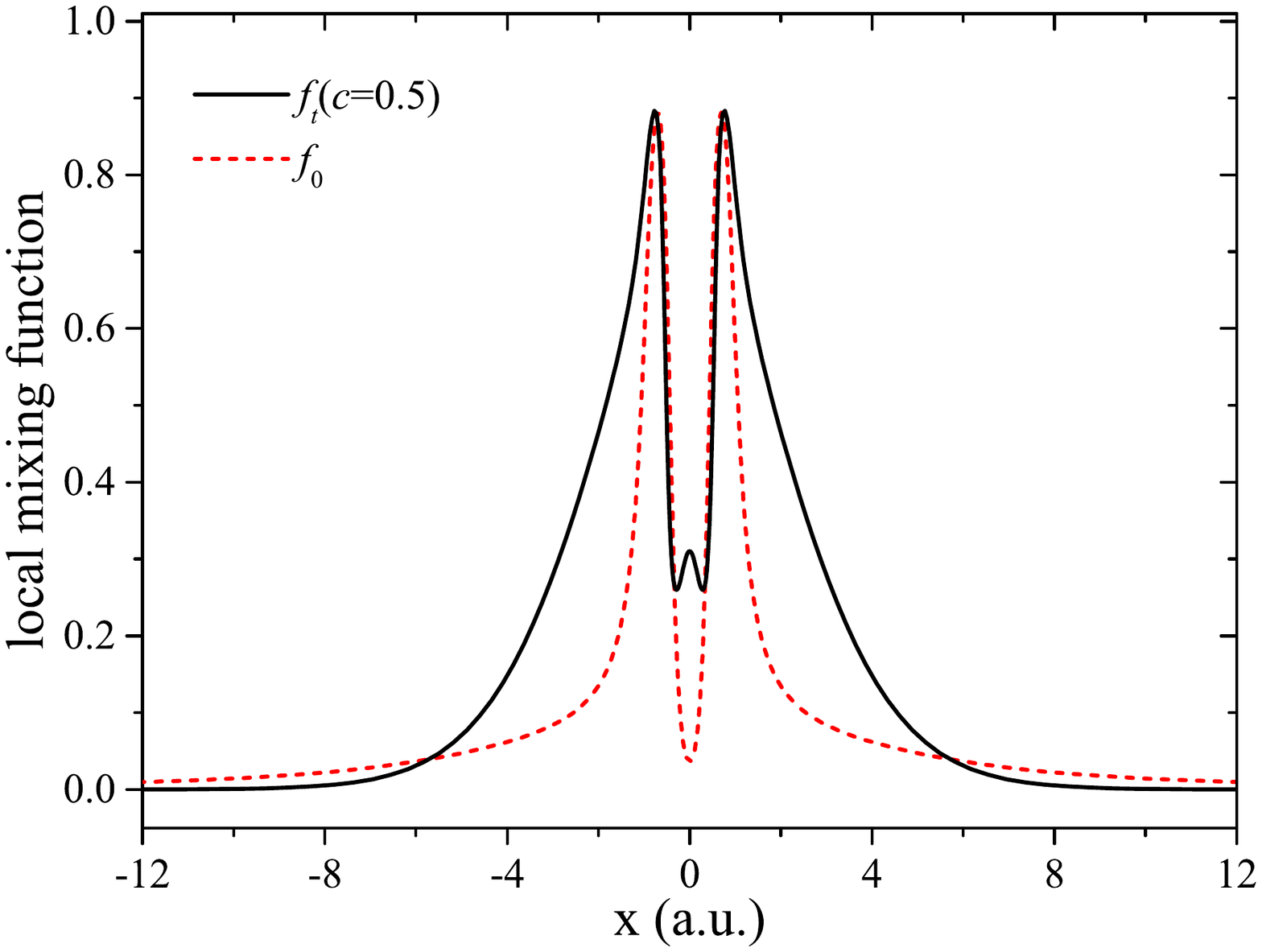}\\
    \caption{Comparison of $f_{0}\rr$ and $f_{t}\rr$ with $c=0.5$ for the C atom along the $x$-axis.}\label{fig.f_comp}
\end{figure}

However, a different picture occurs when evaluating $f_0\rr=1-\frac{\tau_\mathrm{W}\rr}{\tau\rr}$. This function decays much more slowly than $f_t\rr$ with finite $c$, as Fig~(\ref{fig.f_comp}) shows for the carbon atom. 
Consequently, not all terms in the functional derivative originating from $f_0\rr$ vanish individually and more detailed investigations are necessary. 

Defining $K\rr=n\rr(e_\mathrm{x}^{\mathrm{sl}}\rr-e_\mathrm{x}^{\mathrm{ex}}\rr)$, the functional derivative in this case reads
\begin{eqnarray}
 u^{\mathrm{c-nl}}_{i\s}\rr &=& - \frac{f_0\rr}{2}\,u^{\mathrm{exx}}_{i\s}\rr +f_0\rr  v_{\mathrm{x},\s}^{\mathrm{LSDA}} \rr \notag\\
 &+&\frac{1}{2\varphi^*_{i\s}\rr}\sum_{j=1}^{N_{\s}} \varphi^*_{j\s}\rr\notag \int f_0\rrp\frac{\varphi_{i\s}^*\rrp\varphi_{j\s}\rrp}{|\mathbf{r}-\mathbf{r}'|} \,\mathrm d^3r' \notag 
 \notag\\ &-&    \frac{1}{2\varphi^*_{i\s}\rr} \left[\left(\nabla^2\varphi^*_{i\s}\rr\right)\frac{\delta f_0\rr}{\delta \tau\rr}K\rr\right.\notag\\
&+&\left.\nabla\varphi^*_{i\s}\rr\cdot \nabla \left(\frac{\delta f_0\rr}{\delta \tau\rr}K\rr  \right)\right] 
\notag\\
&-&\frac{1}{2n^{\frac{1}{2}}\rr}\left[\left(\nabla^2n^{\frac{1}{2}} \rr\right)\frac{\delta f_0\rr}{\delta \tau_\mathrm{W}\rr}K\rr  \right. \notag\\
& +&\left. \nabla n^{\frac{1}{2}}\rr \cdot \nabla \left(\frac{\delta f_0\rr}{\delta \tau_\mathrm{W}\rr}K\rr\right) \right],  \label{eq.uiso_f0c_omplete}
\end{eqnarray}
with $\frac{\delta f_0\rr}{\delta \tau\rr} = \frac{\tau_\mathrm{W}\rrp}{\tau^2\rrp}= -\frac{\delta f_0\rr}{\delta \tau_\mathrm{W}\rr} \frac{\tau_\mathrm{W}\rr}{\tau\rr}$. Therefore, both $\frac{\delta f_0\rr}{\delta \tau\rr}$ and  $\frac{\delta f_0\rr}{\delta \tau_\mathrm{W}\rr}$  
reach the same absolute value in the asymptotic limit, but show opposite signs. 
Now, we have to distinguish between the spin channels: If one looks at $u^{\mathrm{c-nl}}_{N_{\s_\mathrm{ho}}\s_\mathrm{ho}}\rr$ in the spin channel that has the global HOMO, i.e. $n\rr\sim |\varphi_{N_{\s_\mathrm{ho}}\s_\mathrm{ho}}\rr|^2$, then one can
see from Eq.~(\ref{eq.uiso_f0c_omplete}) that the fourth and fifth term are equivalent in the asymptotic limit except for the sign. Therefore, they cancel each other and, since 
the first and second term decay fast enough, only the third term remains, leading to the limit of $-\gamma_{\s_\mathrm{ho}}/r$. In the other spin channel however, the fourth and fifth term do not cancel anymore, since the density is still dominated by $\varphi_{N_{\s_\mathrm{ho}}\s_\mathrm{ho}}\rr$, whereas the fourth term
features $\varphi_{N_{\bar{\s}_\mathrm{ho}}\bar{\s}_\mathrm{ho}}\rr$. Therefore,
in the other spin channel yet another asymptotic limit is obtained, again strictly following from the evaluation of the functional derivative.

This feature can be corrected by using a spin-polarized ansatz with an indicator function that is a spin-polarized LMF of the form $g_\s\rr=\frac{\tau_{\mathrm{W}\s}\rr}{\tau_\s\rr} $, with $\tau_{\mathrm{W}\s}\rr$ and $\tau_{\s}\rr$ being the kinetic energy spin densities. In this case,
a functional derivative that does not feature the total density $n\rr$ follows and therefore the aforementioned effect does not occur. However, since for the spin channel $\s_\mathrm{ho}$ all derivations 
made are valid independently of the form of the LMF and since this spin channel features the physical meanigful quantity $-\eps_\mathrm{ho}$, it suffices for this work to consider the more simple LMFs instead of their spin-polarized counterparts.

\footnotesize{
\bibliography{bibliography,skrefs} 
\bibliographystyle{rsc} 
} 

\end{document}